\documentclass[aps,twocolumn,amsmath,amssymb,showpacs,prb,floatfix]{revtex4-1}
\usepackage{amsmath}
\usepackage{graphicx}
\usepackage{epsfig}
\newlength{\upit}\upit=0.1truein

\newcommand{\ltappr}{{{\lower4pthbox{$<$} } \atop \widetilde{ \ \ \ }}}
\newlength{\bxwidth}\bxwidth=1.5 truein

\newcommand{\dg}{^{\dagger }}

\newcommand{\rarrow}{\rightarrow}
\newcommand \bea {\begin{eqnarray} }
\newcommand \eea {\end{eqnarray}}
\newcommand{\bk}{{\bf{k}}}
\newcommand{\ba}{{\bf{a}}}
\newcommand{\bq}{{\bf{q}}}

\newcommand{\br}{{\bf{r}}}

\newcommand{\priro}{Pr$_2$Ir$_2$O$_7$ }
\newcommand{\prirop}{Pr$_2$Ir$_2$O$_7$}

\newlength{\figwidth}
\newlength{\shift}
\newcommand{\fg}[3]
{
\begin{figure}[ht]

\vspace*{-0cm}
\[
\includegraphics[width=\figwidth]{#1}
\]
\vskip -0.2cm
\caption{\label{#2}
\small#3
}
\end{figure}}

\begin{document}

\title{Chiral RKKY interaction in Pr$_2$Ir$_2$O$_7$} 

\author{Rebecca Flint and T. Senthil} 
\affiliation{ 
Department of Physics, Massachusetts Institute of Technology, Cambridge, Massachusetts 02139,
U.S.A.  }
 
\begin{abstract} 
Motivated by the potential chiral spin liquid in the metallic spin ice \prirop, we consider how such a chiral state might be selected from the spin ice manifold.  We propose that chiral fluctuations of the conducting Ir moments promote ferro-chiral couplings between the local Pr moments, as a chiral analogue of the magnetic RKKY effect.  \priro provides an ideal setting to explore such a \emph{chiral RKKY effect}, given the inherent chirality of the spin-ice manifold.  We use a slave-rotor calculation on the pyrochlore lattice to estimate the sign and magnitude of the chiral coupling, and find it can easily explain the 1.5K transition to a ferro-chiral state.
\end{abstract} 
 
\maketitle

\section{Introduction}

\priro sits at the intersection of two recent fields of high interest in condensed matter physics: spin ice and iridate physics\cite{Nakatsuji06,Machida10}.  Pr and Ir sit on interpenetrating pyrochlore lattices of corner sharing tetrahedra, where the Pr ions are Ising local moments while the Ir form a correlated conduction band.  In such a heavy fermion material, one expects magnetic order or a heavy Fermi liquid, but there is no sign of either.  Instead, below $1.5$K there is an anomalous Hall effect that persists in the absence of both magnetic field and any observable magnetization. This hysteretic signal implies the onset of an unknown phase, originally proposed as a chiral spin liquid\cite{Machida10}.

Pr$^{3+}$ is a 4f$^2$ ion with a $J=4$ magnetic non-Kramers doublet $|\Gamma_3 \pm \rangle \approx |J_z = \pm 4\rangle$ ground state\cite{Machida05}.  The Pr moments are therefore perfectly Ising and point into and out of the tetrahedra along the local $\langle 111 \rangle$ axis, while the in-plane components of the doublet are strictly quadrupolar.  The RKKY interaction gives a ferromagnetic Pr-Pr coupling\cite{Machida10}, which means the Pr sublattice realizes the spin ice Hamiltonian.  Spin ice is a degenerate ground state on the pyrochlore lattices containing all possible arrangements of ``2 in - 2 out'' tetrahedra\cite{Bramwell01} (see Fig. 1A).  Indeed, spin-ice correlations are seen in the magnetization, where a meta-magnetic transition to the ``3 in - 1 out'' state is seen for fields along the $[111]$ direction, but not $[100]$ and $[110]$, where the field simply aligns ``2 in - 2 out'' tetrahedra\cite{Machida10}. The resistivity minimum at 25K is also consistent with the development of spin-ice\cite{Udagawa12}. The data is strongly suggestive of spin-ice correlations, but not conclusive in the way pinch-points in neutron scattering would be\cite{Isakov04}.  However, spin-ice is clearly not the whole story, as \priro has no extensive ground state entropy, implying further correlations.

The Ir$^{4+}$ configuration is 5d$^5$, equivalent to one hole in the t$_{2g}$ orbital.  Strong spin-orbit coupling is predicted to convert this configuration to a half-filled $J_{\rm eff} = 1/2$ doublet\cite{Kim08}.  As there are 4 Ir/unit cell, semi-metallic behavior is expected.  However, due to the narrowed $J_{\rm eff} = 1/2$ band-width, the smaller 5d U is capable of inducing Mott insulating behavior, as seen in Sr$_2$IrO$_4$\cite{Kim08} and Na$_2$IrO$_3$\cite{Singh10}.
\priro is part of a family of pyrochlore iridates, R$_2$Ir$_2$O$_7$ that undergo a metal-insulator transition (MIT) as a function of the rare-earth element\cite{Yanagashima01,Matsuhira07,Matsuhira11,Zhao11}. Numerical studies of the pyrochlore iridates predict not only metallic and Mott insulating states, but an intermediary Weyl semi-metal\cite{Wan11}.  While R=Pr is metallic, R=Nd-Ho have MITs at temperatures increasing from 36K to 150K.  Pressure experiments on Nd$_2$Ir$_2$O$_7$\cite{Sakata11} and Eu$_2$Ir$_2$O$_7$\cite{Tafti11} have been able to suppress the insulating phase, indicating that \priro is quite close to a MIT induced by increasing interaction strengths.  The exact nature of this transition is unclear; the second order nature suggests that it has a large Slater component, but analogy with other iridates suggests that it likely has some Mott character.

The Pr moments and Ir electrons are coupled by the Kondo effect, but the non-Kramers nature means that valence fluctuations are to excited Kramers \emph{doublets}, generating a two-channel Kondo effect instead of the usual one-channel effect.  Here, the symmetry of the two channels is protected by time-reversal symmetry.  While the ordinary Kondo effect leads to heavy Fermi liquid formation, the two-channel version is critical and cannot form a Fermi liquid without a phase transition\cite{2CK,Hastatic}.  The diverging Sommerfeld coefficient, $\gamma(T) = C(T)/T$, the partial quenching of the Pr moment seen in the susceptibility and the resistivity minimum are all consistent with two-channel Kondo physics\cite{2CK,Machida07}.  While two-channel Kondo physics can lead to broken symmetry states, as proposed for URu$_2$Si$_2$[\onlinecite{Hastatic}], the main signature of these hastatic states is the development of a heavy Fermi liquid at a mean-field-like phase transition.  In \prirop, $\chi$ and $\gamma$ continue to diverge and there is no sharp phase transition in the thermodynamic quantities.

The absence of thermodynamic anomalies is also the strongest evidence against magnetic order.  It is supported by the absence of spin-freezing in $\mu$SR data\cite{MacLaughlin09}, although, as the muons modify the Pr environment, neutron or x-ray experiments would be more conclusive. 

There is, however, a clear signature in the Hall effect, which shows hysteresis in field and develops a zero-field contribution sharply at 1.5K, while the susceptibility does not develop hysteresis until much lower temperatures\cite{Machida10}.  The experimentalists proposed that this large Hall effect is due to a macroscopic scalar spin chirality.  The scalar spin chirality, $\kappa_{ijk} = \vec{S}_i \cdot \vec{S}_j \times \vec{S}_k$ is the solid angle between three spins on a triangular plaquette - as such it is only nonzero when the three spins are neither collinear nor coplanar.  In metallic systems, the chirality's identity as a solid angle means that conduction electrons coupled to the local moments acquire a Berry phase as they move through the lattice.  This Berry phase acts as a local magnetic flux and contributes to the anomalous Hall effect, $\rho_{xy} = R_0 B + R_M M + R_\kappa \kappa$, even in the absence of both magnetic field and magnetization\cite{Nagaosa10}.  This chiral contribution can develop in magnetically ordered states, like Nd$_2$Mo$_2$O$_7$[\onlinecite{ndmo}], but does not require magnetic order.  Several recent theories for \priro have proposed that $R_M$ is anomalously large, while $\kappa$ is irrelevant\cite{udegawa12,moon12}. However, as $B=0$ and $\vec{m}$ is unobservably small, we take the experimentalists' point of view that the chirality is the main event.  This picture can be checked - the Kerr effect can in principle measure the broken time-reversal, and resonant inelastic x-ray scattering (RIXS) should be able to detect spin-chirality directly\cite{Ko11}.

The Ising pyrochlore lattice has a lot of inherent spin chirality due to the non-coplanar Ising axes, which give every triangular plaquette a non-zero chirality.  But is it possible to realize chirality without magnetization?  While the chirality of the tetrahedron is always proportional to its magnetization, we can construct superpositions of states with large chirality and zero magnetization where the chirality originates from the hexagons of the kagom\'{e} planes.  However, a simple symmetry analysis actually reveals that any chiral state will automatically also be magnetic.  

In three dimensions, the scalar chirality acquires a direction given by the normal to the plaquette.  The point group of the pyrochlore lattice is $\rm{O}_h$, and both chirality and magnetization transform as the three-dimensional $\Gamma_4$ irrep.  Therefore a linear coupling between the two, $\gamma \vec{m} \cdot \vec{\kappa}$ is expected.  This problem is peculiar to the pyrochlore lattice, as generally the chirality and magnetization transform differently.  One might expect that a different time-reversal symmetry breaking order parameter, for example, one with $\Gamma_5$ symmetry could explain the lack of magnetization, however the existence of an anomalous Hall effect requires a $\Gamma_4$ order parameter, as the conduction electron Berry phase transforms as $\Gamma_4$.  Experimentally, it appears that there is a large chirality without a large magnetization, so it is an important open question why the symmetry allowed linear coupling appears to be so small in \prirop.

\fg{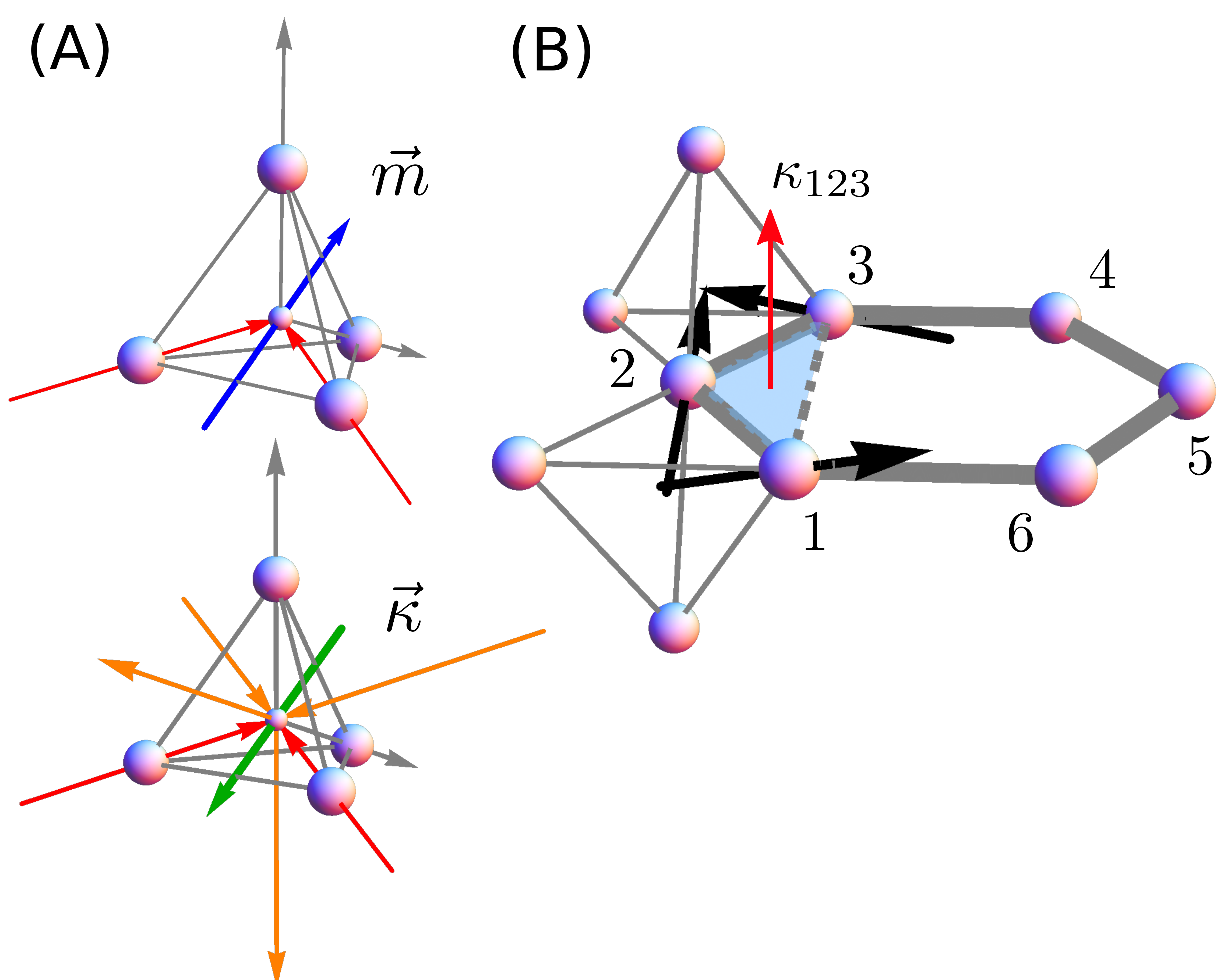}{chirality}{(Color online) (A) For any given tetrahedra, the net chirality (a vector sum over the faces) will be proportional to the net magnetization. (B) Thus for the system to exhibit net magnetization without net chirality, the chirality must originate from independent triangles, like the isosceles triangles that make up the hexagons of the kagom\'{e} planes, whose chirality will have an associated direction given by the normal to the triangular plaquette.}

In a 2D system, the chirality is an Ising quantity, and as such a chirality transition should be in the Ising (O(1)) universality class, with a diverging specific heat.  However, in cubic symmetry, the chirality can point along eight symmetry equivalent directions, meaning that it should instead be described as an O(3) vector order parameter with strong pinning to those directions, which gives a kink in the specific heat.  The specific heat is presumably broadened by disorder in the real material, leading to the observed broad peak.

As the spin-ice manifold contains chiral as well as non-chiral states, we propose that some interaction favors these chiral states, leading to the development of a correlated state with a large chirality.  This state is essentially a chiral spin liquid, except for the small parasitic magnetic moment guaranteed by symmetry - we will call this state a CSL*.  A chiral spin liquid is a magnetic state that breaks time-reversal symmetry, but no others; specifically, it has no long range magnetic order\cite{Kalmeyer87}, but does have a uniform $\langle \kappa \rangle$. Theorists originally discussed \emph{quantum} chiral spin liquids\cite{Kalmeyer87,Wen89}, but it is also possible to have finite temperature \emph{classical} chiral spin liquids, which would occur above low temperature magnetically ordered chiral states.  The initial theoretical proposals were in insulating two-dimensional Heisenberg magnets\cite{Wen89}, so it is quite interesting to find a strongly related state in a metallic three-dimensional Ising magnet.  This CSL* state is not differentiated from a ferromagnetic state by symmetry, although a topological phase transition is possible.  In order to understand the CSL*, we must first understand its Hamiltonian.  The natural question to ask is: what kind of perturbation can one add to the magnetic Hamiltonian to select this chiral state out of the spin-ice manifold?  There are two generic possibilities for \prirop: the RKKY interaction will generate further neighbor interactions or quantum exchange terms, which are enhanced over those in canonical spin ice by both the smaller Pr moment and the Kondo effect, can kinetically favor certain configurations. 

The magnetic interactions between Pr ions originates from the RKKY interaction\cite{RKKY}, where the non-Kramers nature of the Pr moments ensures that the magnetic RKKY interaction is purely Ising-like, as the in-plane moment of the $\Gamma_3$ doublet is quadrupolar in nature; we are neglecting the quadrupolar exchange. The RKKY interaction is long-ranged, decaying as $1/r^3$ just like the dipolar interaction.  In insulating spin ice, the dipolar interactions delicately balance to preserve the degenerate spin ice state\cite{Bramwell01}, however, the RKKY interaction does not generically do so\cite{Ikeda08}.  While the further neighbor interactions can select states out of the spin-ice manifold, there is no natural way to select a non-magnetic chiral state - even magnetically ordered states with chirality, but no magnetization require large unit cells.  Furthermore, in three dimensions we do not expect the magnetic and chiral ordering temperatures to split.  So further neighbor RKKY interactions do not provide a satisfying explanation for a CSL*.

What about non-Ising exchange terms, which have been predicted to generate spin liquid states\cite{Hermele04}?  The above picture is purely classical, which works well for the large moments in insulating spin-ice, $\mu_{\rm eff} \sim 10 \mu_B$, but here the moments, $\mu_{\rm eff} \sim 3 \mu_B$ are much smaller, and quantum effects likely become important.  Quantum effects can be treated by perturbatively introducing quantum exchange terms, $\sum_{\langle ij\rangle} J_\perp S_{i+} S_{j-} + {\rm H.c}$; the specific origin is addressed by Chen and Hermele\cite{Chen12}.  This problem was originally considered by Hermele, Fisher and Balents\cite{Hermele04}, who suggested such a Hamiltonian might lead to a $U(1)$ quantum spin liquid.  The main idea is that degenerate perturbation theory favors states that can be connected by flipping every spin in a loop.  The first such term comes from third order perturbation theory and involves flipping every spin in a hexagon, favoring ``flippable hexagons,'' where ``in'' and ``out'' spins alternate, as shown in Fig 2.  As these hexagons have zero net chirality, the $U(1)$ spin liquid is not a chiral spin liquid, although it may be related to materials like Yb$_2$Ti$_2$O$_7$\cite{TbTi,TbTiTheory}.  By inspection, the sixth order loops also do not favor chiral configurations, so the chiral ordering temperature is at best $O(J_\perp^9/(2 J_z)^8)$.  Even taking $J_\perp = J_z$, this term is $.004 J_z$, three orders of magnitude too small. So quantum fluctuations cannot generate a chiral spin liquid state at $T_H = 1.5$K.

In the next section, we explain how local moment chirality interactions can be mediated by chirality fluctuations of the conduction electrons, giving rise to a chiral RKKY effect in section II.  Then in section III, we estimate the magnitude and sign of the chiral RKKY effect for Ir electrons on the pyrochlore lattice.  Finally, in section IV, we discuss the properties of this kind of chiral spin liquid and how it may be tested.

\fg{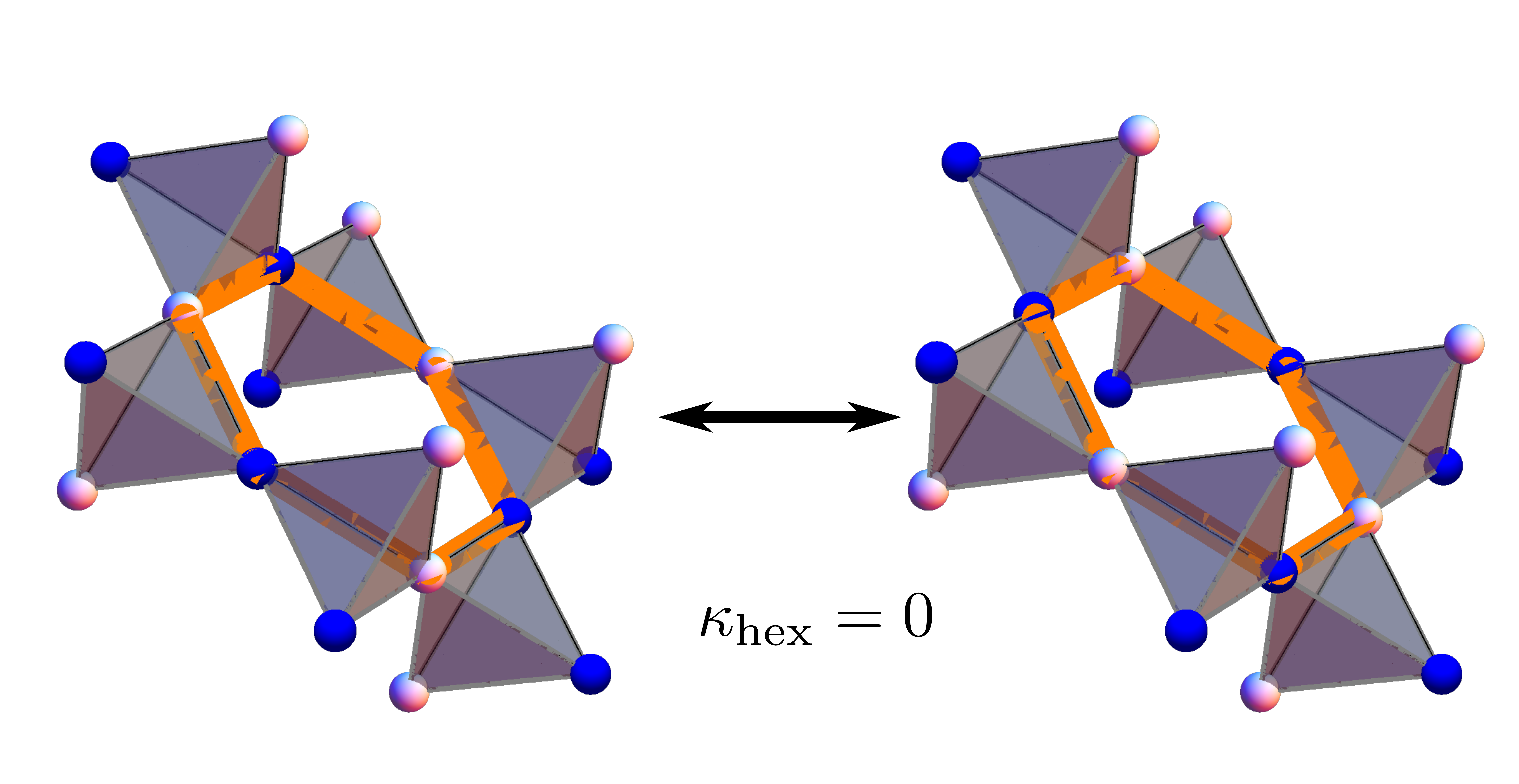}{magnetism}{(Color online) Hexagons which contain alternating ``in'' and ``out'' spins contain local modes in which every spin around the hexagon is flipped, resulting in another state in the spin-ice manifold.  These ``flippable'' hexagons have zero chirality.  Here, we indicate ``in'' spins with blue (darker) sphere and ``out'' with white (lighter) spheres, and have defined ``in'' and ``out'' self-consistently around the hexagon by picking the axes pointing into the alternating tetrahedra pointing out of the plane.}
% (B) The ordered state favored by ferromagnetic second-neighbor interactions.  One out of every three hexagons has the same net non-zero chirality and net magnetization, leading to a ferromagnetic, ferrochiral state.  (C) The ordered state favored by ferromagnetic third-neighbor interactions.  Each hexagon carries the same chirality and magnetization, again leading to a ferromagnetic, ferrochiral state.}

\section{Chiral RKKY effect}

As magnetic interactions are unlikely to generate the $T_H = 1.5$K transition in \prirop, we must look elsewhere for the origin of the chiral state: to the conduction electrons. Just as conduction electrons carrying spin mediate the magnetic RKKY effect between local moments\cite{RKKY}, interactions between the local moment chiralities can be mediated by chirality fluctuations of the conduction electrons, $\kappa_c(\br)$.  Generically, this effect will be captured by the Hamiltonian,
\begin{equation}
H = \sum_{\br,\br'} \kappa_c(\br)\chi_\kappa^{-1}(\br-\br')\kappa_c(\br') + \sum_{\langle \br,j\rangle} J_{cf} \kappa_c(\br)\kappa_{f}(j),
\end{equation}
where $\kappa_c(\br)$ represents the conduction electron chirality, $\kappa_{f}(j)$ represents the local moment chirality of a plaquette centered at $j$, $\chi_\kappa(\br-\br')$ is the conduction electron chiral susceptibility, and $J_{cf}$  is the coupling between the local moment and conduction electron chirality that originates from the Kondo interaction, $J_{cf} \propto J_K^3/\epsilon_F^2$.  The direct local moment chiral coupling, $\sum_{ij} J_{ff}(i-j) \kappa_f(i)\kappa_f(j)$ is likely to be too small, however we shall show that a large coupling can be generated by integrating out the conduction electrons:
\begin{equation}
J_{ff}(\br-\br') = -|J_{cf}|^2 \chi_\kappa(\br-\br'),
\end{equation}
in exact analogy with the magnetic RKKY effect.  Hence, we dub it the \emph{chiral RKKY effect}.
Such a term will be generated directly from integrating out the conduction electrons directly from the Kondo Hamiltonian, as a sixth order term of order $J_K^6/\epsilon_F^5$; however, we prefer this second order picture, both for the better understanding of the physical mechanism and also for the relative simplicity of the calculation.  The magnitude of this chiral RKKY effect can be roughly estimated from experimental parameters.  If we take the single band estimate of $n = 4.12\times 10^{21}$cm$^{-3}$[\onlinecite{Machida10}] and the bare electron mass, the Fermi energy is $\epsilon_F \approx 10^4$K. Then taking $J_{RKKY} = \theta_{CW} = 20K$[\onlinecite{Machida10}],  these parameters lead to a Kondo coupling, $J_K = 310$K and a chiral RKKY effect on the order of $J_K^6/\epsilon_F^5 \sim 0.1$K.  This rough estimate gives a coupling only an order of magnitude smaller than the actual ordering temperature.  However, we can make a better estimate using equation (2), and we find a ferrochiral coupling one to two orders of magnitude larger than this estimate.  There are two straightforward ways to estimate this coupling, and we begin with a simple classical picture in the limit of infinite Kondo coupling to illustrate the general principle, before going on to a more microscopically motivated slave-rotor approach.  

\section{Estimate \# 1: classical moments}

A simple upper bound to the chiral coupling can be found by taking the limit of infinite Kondo coupling, $J_K \rarrow \infty$ and assuming the Pr moments to be classical.  The conduction electron spins then track the local moments perfectly as they move around the lattice, seeing the local moments as a background gauge field, $\mathcal{A}(\br)$\cite{backgroundgauge}. The local moment chirality is given by $\frac{A}{\hbar}\nabla \times \mathcal{A}(\br)$, where $\frac{A}{\hbar}$ ensures the chirality is dimensionless, and $A$ is the area of a plaquette. Here, we take the conduction electrons to have a parabolic dispersion, $\epsilon_k = k^2/2m -\mu$, and their action in the background gauge field is,
\begin{equation}
S[c,\mathcal{A}] = k_B T \sum_{i\omega_n} \int_\bk c\dg_\bk\left[ i\omega_n - \frac{(\hbar \bk-\mathcal{A})^2}{2m} -\mu\right] c_\bk.
\end{equation}
After integrating out the conduction electrons, the action becomes
\begin{equation}
S[\mathcal{A}] = \int_\bq \mathcal{A}(\bq)\Pi_C(\bq)\mathcal{A}(-\bq),
\end{equation}
where $\Pi_C(\bq) = 1/\langle \mathcal{A}(\bq) \mathcal{A}(-\bq)\rangle^T = -\langle J_C(\bq) J_C(-\bq)\rangle^T$ is the transverse current-current correlator for the conduction electrons, and we have already taken the static limit for simplicity.  This action can be rewritten in terms of the chirality, $\kappa(\bq)$,
\begin{equation}
S[\kappa] =  \frac{\hbar^2}{A^2}\int_\bq \kappa(\bq) \frac{\Pi_c(\bq)}{\bq^2}\kappa(-\bq).
\end{equation}
For a parabolic dispersion, the current-current correlator can be calculated analytically in the zero temperature limit,
\bea
\Pi_C(\bq) & = & 
\frac{2n}{m} + \frac{2\hbar^2 k_B T}{3m^2}\sum_{i\omega_n}\int_\bk \vert\bk + \frac{\bq}{2}\vert^2\frac{f(\epsilon_{\bk+\bq})-f(\epsilon_\bk)}{\epsilon_{\bk}-\epsilon_{\bk+\bq}}\cr
& = & \!\frac{n}{m}\!\left[\frac{1}{2}+\frac{\tilde{q}^2}{2}-\frac{1}{4}\! \left(\frac{1}{\tilde{q}} - 2\tilde{q}+\tilde{q}^3\!\right)\!\log\left\vert \frac{1+\tilde{q}}{1-\tilde{q}}\right\vert\right]\!,
\eea
where we have introduced $\tilde{q} = q/2k_F$ for clarity. 

The chiral coupling constant, $J_\kappa(r)$ can now be calculated
\bea
J_\kappa(r) & = & \frac{\hbar^2\Omega}{(2\pi)^2 A^2} \int_0^\infty dq q^2 \int_{-1}^{1}d \cos \theta \frac{\Pi_C(\bq)}{q^2} e^{i q r \cos \theta}\cr
& = & \frac{\hbar^2\Omega}{2 \pi^2 A^2 r} \int_0^\infty d\tilde{q} \frac{\Pi_C(q) \sin 2 k_F \tilde{q} r}{\tilde{q}}.
\eea
The result is plotted in Figure 3, using $k_F = 2\pi/(1.218 a)$ taken from the Hall effect\cite{Machida10,Ikeda08}, $n = 4.12\times 10^{21} {\rm cm}^{-3}$ and the lattice constant, $a = 10.4\AA$.  The finite $k_F$ leads to oscillatory behavior, so both ferrochiral and anti-ferrochiral interactions are possible, depending on the distance.  For the parameters of \priro, the chiral coupling at $r = a$ is -200K. As we have taken $J_K \rarrow \infty$, this estimate is obviously an upper bound on the real value.

\fg{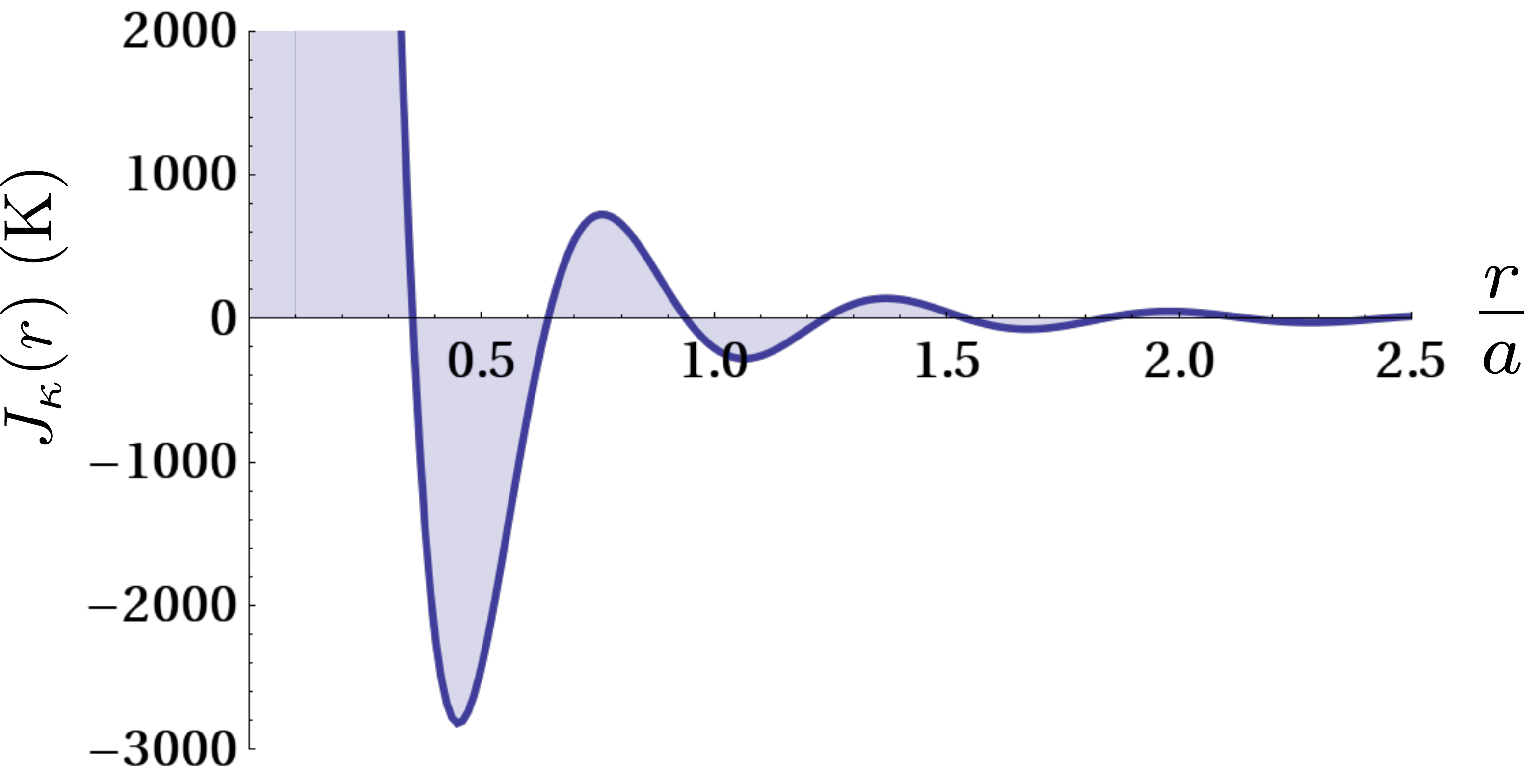}{classical}{The chiral coupling $J_\kappa(r)$ calculated in the classical, $J_K \rarrow \infty$ approximation for a parabolic dispersion.  The y-axis is in Kelvin, for a possible set of parameters for \prirop.  For $r = a = 10.4\AA$, the chiral coupling is -200K.  Just like the magnetic RKKY interaction, the chiral RKKY interaction is oscillatory and decays rapidly.}

\section{Estimate\#2: slave rotors}

To make a more accurate estimate, we return to equation (2), and now attempt to calculate the conduction electron chiral susceptibility, $\chi_\kappa(\br-\br')$. As it is a twelve point correlation function, this calculation initially appears quite tedious.  However, by introducing the slave-rotor mean-field approximation\cite{Florens04}, we can greatly simplify the calculation, and in the process model the approach to the Mott transition.  As electrons approaching a Mott transition have increased chiral fluctuations\cite{Lee92}, we expect the coupling to be enhanced.

Slave-rotors are typically introduced to capture the quantum criticality associated with the Mott transition in the Hubbard model by splitting the conduction electron into a neutral spinon and a charged holon, $c^\dagger_{i\sigma} = f^\dagger_{i\sigma} e^{i \theta_{i}}$[\onlinecite{Florens04}].  This decoupling possesses a $U(1)$ gauge symmetry,
\begin{equation}
{\rm U(1)\; gauge\; symmetry: }\;
\left\{\begin{array}{c}
f^\dagger_{i\sigma} \rightarrow f^\dagger_{i\sigma} e^{i a_i}\\
\theta_i \rightarrow \theta_i - a_i
\end{array}\right.,
\end{equation}
which glues the spinon and holon back together to form the charged electron in the metallic phase.  The metallic phase is captured by the uniform condensation of this rotor, $\langle e^{i\theta_i}\rangle$, while it is uncondensed in the Mott insulating phase and the Mott transition is a 4D XY transition before the coupling to the gauge field is taken into account.
The spin chirality is particularly simple in this approach\cite{Lee92},
\begin{equation}
\nabla \times \ba = \frac{\hbar}{A} \vec{s}_{c1} \cdot \vec{s}_{c2} \times \vec{s}_{c3},
\end{equation}
where $\nabla \times \ba$ is the lattice curl of the gauge field around a triangular plaquette, and the units are fixed by $\frac{\hbar}{A}$ as before.  The chiral susceptibility can then be calculated from the transverse gauge propagator, $\langle \ba(\bq)\ba(-\bq)\rangle^T$,
\bea
\chi_\kappa^{ab}(\bq)& =& \langle \kappa^a_c(\bq) \kappa^b_c(-\bq)\rangle\cr
& = & \frac{A^2}{3\hbar^2} (q^2 \delta_{ab} -q_a q_b)\langle \ba(\bq)\ba(-\bq)\rangle^T.
\eea
As the spinon and holon add in series, the gauge propagator is,
\begin{equation}
\langle \ba(\bq) \ba(-\bq)\rangle^T = \left[\Pi^T_{F}(\bq) + \Pi^T_{X}(\bq)\right]^{-1},
\end{equation}
where $\Pi^T_{F,\theta}(\bq)$ are the transverse current-current correlators for the spinons and rotors, respectively.

We now calculate these correlators within a specific model: electrons on the half-filled Ir pyrochlore lattice, where the strong spin-orbit coupling means the 5d$^5$ Ir configuration is simplified to an isotropic effective $J_{\rm eff} = 1/2$ doublet\cite{Pesin10}. Both oxygen mediated Ir-O-Ir and direct Ir-Ir hopping are expected\cite{Witczak12}, and we take the nearest-neighbor dispersion considered by Witczak-Krempa et al\cite{Witczak12}, where the oxygen mediated hopping has magnitude $t_O$ and the direct hopping is governed by two parameters, $t_\sigma$ and $t_\pi$, with the specific choice $t_\sigma = -t_O$, $t_\pi = \frac{2}{3}t_O$.  This dispersion has no Fermi surface at half-filling, instead having a peculiar quadratic band touching with zero density of states at $E_F$.  We add a small next-nearest-neighbor oxygen mediated hopping, $t_2$ to convert the quadratic band touching into a semi-metallic dispersion with small electron and hole pockets. The strong spin orbit coupling means that the Ir-Ir hopping is spin-dependent, leading to a complex hopping matrix $\Gamma_{\sigma\sigma'}^{(ia)(jb)}$, where $a,b \in \{1-4\}$ label the four sites per unit cell. 
The Hubbard model for this system is then,
\bea
H & = & -\mu \sum_{ia \sigma} c^\dagger_{ia\sigma} c_{ia\sigma} + \frac{U}{2} \sum_{ia}(\sum_\sigma c^
\dagger_{ia\sigma} c_{ia\sigma} -\frac{N}{2})^2 \cr
& & + \!\! \sum_{iajb\sigma\sigma'} t_{ij}\Gamma_{\sigma\sigma'}^{(ia)(jb)} c^\dagger_{ia\sigma}c_{jb\sigma'}.
\eea
As the nearest neighbor hopping has several components, hidden within $\Gamma_{\sigma\sigma'}^{(ia)(jb)}$, $t_1 \equiv t_{\langle ij\rangle}$ is actually just $t_O$.
Here the conduction electron density is kept fixed to half-filling by $\mu$, and the proximity to the Mott transition is tuned by changing the interaction strength, $U$.  We introduce the slave-rotor decoupling, $c^\dagger_{i a\sigma} = f^\dagger_{ia\sigma} e^{i \theta_{ia}}$, where from now on we will work in terms of $X_{ia} \equiv e^{i\theta_{ia}}$.  The hopping term is now a quartic term, and we introduce four Hubbard-Stratonovich fields to decouple it:
\bea
Q_F^{(1,2)} & \propto & \langle X_{ia} X_{jb}^*\rangle\vert_{\langle ij \rangle, \langle\langle ij \rangle\rangle}\cr
Q_X^{(1,2)} & \propto & \sum_{\sigma,\sigma'} \Gamma_{\sigma \sigma'}^{(ia)(jb)} \langle f^\dagger_{ia \sigma} f_{jb\sigma'}\rangle\vert_{\langle ij \rangle, \langle\langle ij \rangle\rangle},
\eea
where $\langle ij \rangle$ and $\langle\langle ij \rangle\rangle$ are really a shorthand restricting $(ia)$ and $(jb)$ to be nearest or next-nearest neighbors, respectively.  In other words, we are assuming the $Q_{F,B}^{(ia)(jb)}$s take one value, $Q^1$ for nearest neighbors, another, $Q^2$ for next-nearest neighbors, and are zero otherwise.
This decoupling leads to quadratic actions for both the fermions and bosons,
\begin{widetext}
\bea
S_F & = &\int_0^\beta d\tau  \sum_{ia\sigma} f^\dagger_{ia\sigma} (\partial_\tau +\mu) f_{ia\sigma} - t_1 Q_F^{(1)}\!\!\!\! \sum_{\langle ij \rangle,ab,\sigma\sigma}\!\!\!\! \Gamma_{\sigma\sigma'}^{(ia)(jb)} f^\dagger_{ia\sigma} f_{jb\sigma'} - t_2 Q_F^{(2)}\!\!\!\! \sum_{\langle\langle ij \rangle\rangle,ab,\sigma\sigma}\!\!\!\! \Gamma_{\sigma\sigma'}^{(ia)(jb)} f^\dagger_{ia\sigma} f_{jb\sigma'}\cr
S_X & = & \int_0^\beta d\tau \sum_{ia} \frac{\left|\partial_\tau X_{ia}\right|^2}{U} + \lambda_{ia}\left(|X_{ia}|^2 -1\right) -\frac{2}{3}t_1 Q_X^{(1)}\!\! \sum_{\langle ij\rangle, ab}\!\! X_{ia}^* X_{jb}-t_2 Q_X^{(2)}\!\!\!\! \sum_{\langle\langle ij\rangle\rangle, ab}\!\!\!\! X_{ia}^* X_{jb},
\eea
\end{widetext}
where $\lambda_{ia}$ is a Lagrange multiplier enforcing the phase nature of the slave rotors that we will take to be uniform, $\lambda_{ia} = \lambda$, and we have rescaled $U \rarrow U/2$ to preserve the atomic limit\cite{Florens04}.  Note that while the fermions still have the complex hopping matrix of the original conduction electrons, the bosons behave like spinless fermions.  
The nearest neighbor hopping element is then given by the trace of 
$t_O \Gamma_{\sigma \sigma'}^{(ia)(jb)} = t_O+t_\sigma+t_\pi = \frac{2}{3} t_1$.
As the slave rotor technique is fairly standard, and has already been applied to the pyrochlore lattice, with different dispersions\cite{Pesin10,Kargarian11}, we have skipped several intermediate steps. 
The fermionic and bosonic bands are shown in Figure 4 A and B, respectively.  The Green's functions are 
\bea
G^X_{kn}(i\nu_m) & = & X_0^2 \delta(k)\delta_{n,0} \delta(i\nu_m) +\frac{1}{\nu_m^2/U + \lambda + \zeta_{kn}}\cr
G^F_{km}(i\omega_n) & = & \frac{1}{i\omega_n - \epsilon_{km}},
\eea
where $\epsilon_{km}$ are the eight fermion bands and $\zeta_{kn}$ are the four bosonic bands.  We have allowed the bosons to condense in the lowest band with the uniform amplitude $X_0 \equiv \langle X_{ia}\rangle$, which implies that the Lagrange multiplier $\lambda = 4 t_1 Q_X^{(1)} + 12 t_2 Q_X^{(2)}$
% used factor of 2/3
 is fixed to the bottom of the bosonic bands.  We will concentrate on the mean-field theory of the metallic phase and quantum critical point at zero temperature, where $X_0 \in [0,1]$ quantifies the distance from the quantum critical point.

\figwidth=8.1cm
\fg{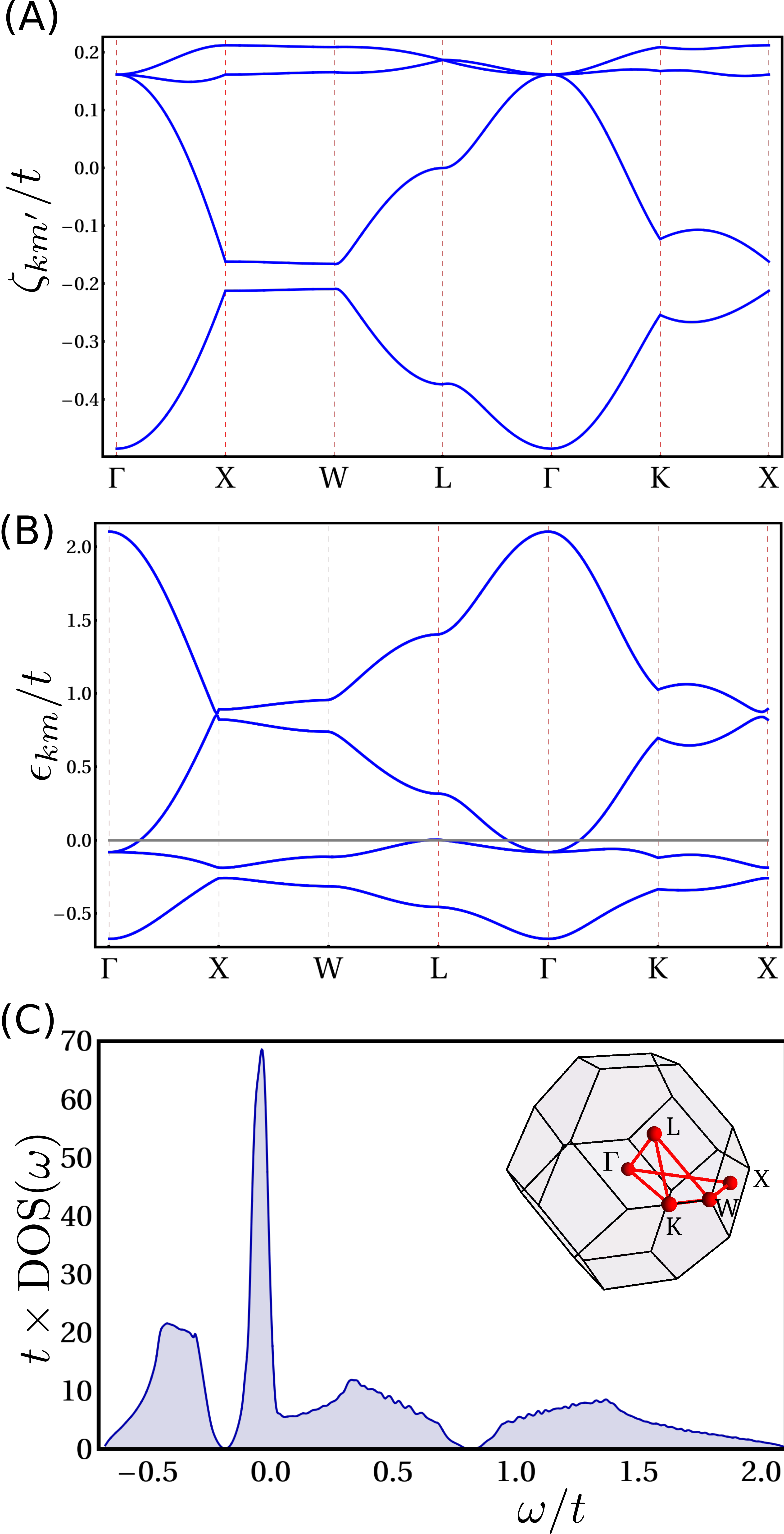}{pyrochlore}{(Color online) Bands on the pyrochlore lattice. (A) The bosonic bands plotted through the high symmetry points of the pyrochlore Brillouin zone, for $t_2 = -.03 t$, $Q_{X}^{(1)} = .14$, $Q_{X}^{(1)} = .20$, where $t$ is the nearest neighbor oxygen-mediated hopping magnitude.  (B) The fermionic bands with $t_\sigma = -t$, $t_\pi = 2/3 t$, and $t_2 = -.03 t$. These are plotted for $U = 0$, and where $\mu = -.31 t$ has been included so that $E_F$ is at zero when the Ir sites are half-filled. (C) The fermionic density of states. (Inset) The pyrochlore Brillouin zone.}

The remaining parameters ($X_0^2, Q_F^{(1,2)}, Q_X^{(1,2)}, \mu$) can be determined from the six mean field equations,
\bea
|X_{ia}|^2 & = & 1 \cr
& = & X_0^2+\frac{1}{\mathcal{N}_s}\sum_{kn} \sqrt{\frac{U}{4(\lambda + \zeta_{kn})}} \cr
Q_F^{(1,2)} & = & \frac{t_{1,2}}{D_{1,2}}\langle X_{ia} X_{jb}^*\rangle\vert_{\langle ij \rangle, \langle\langle ij \rangle\rangle}\cr
& = & X_0^2-\frac{1}{D_{1,2}\mathcal{N}_s} \sum_{kn} \zeta_{kn}^{(1,2)} \sqrt{\frac{U}{4(\lambda + \zeta_{kn})}}\cr
Q_X^{(1,2)} & = & \frac{3t_{1,2}}{2D_{1,2}}\sum_{\sigma,\sigma'} \Gamma_{\sigma \sigma'}^{(ia)(jb)} \langle f^\dagger_{ia \sigma} f_{jb\sigma'}\rangle\vert_{\langle ij \rangle, \langle\langle ij \rangle\rangle}\cr
& = & -\frac{3}{2D_{1,2} \mathcal{N}_s}\sum_{km} \epsilon_{km}^{(1,2)} \theta(\mu-\epsilon_{km})\cr
1 & = & \frac{1}{\mathcal{N}_s}\sum_{km} \theta(\mu-\epsilon_{km}),
\eea
where $D_1 = 6 t_1$ and $D_2 = 12 t_2$ are determined by the relevant number of neighbors, and $\mathcal{N}_s$ is the total number of sites. For $t_2 \neq 0$, these must be solved self-consistently.  We choose $t_2 = .05 t_1$, and find $U_c = 1.35 t_1$.  $X_0^2 = 1-\sqrt{\frac{U}{U_c}}$, while $Q_X^{(1,2)} = .14,.20$ are independent of $U$.  The $U$ dependence of the renormalized bandwidths, $Q_{X,F}^{(1,2)} = 1-\sqrt{\frac{U}{\tilde{U}_{c(1,2)}}}$, where $\tilde{U}_{c(1,2)} = 1.5 t_1, 1.6 t_1$; it is important to note that while the quasiparticle weight, $Z = X_0^2$ goes to zero at the quantum critical point (QCP), the bandwidths decrease with increasing $U$, but remain finite at the QCP.

With the mean field parameters in hand, we can now calculate the current-current correlators, where the cubic symmetry implies $\Pi_{F,X}^T(\bq) = 2\Pi_{F,X}^{yy}(\bq = q\hat x)$.
\begin{equation}
\Pi_{F,X}^{yy}(\bq) = -\langle j_{F,X}^y(\bq) j_{F,X}^y(-\bq)\rangle - \langle T_{F,X} \rangle,
\end{equation}
and the paramagnetic currents are given by,
\bea
j_F^y(\bq) & = & T \sum_{i\omega_n} \sum_{km} \frac{1}{2} (v_{km}^y + v_{k+qm}^y) f_{km}\dg f_{k+qm}\cr
j_X^y(\bq) & = & T \sum_{i\nu_n} \sum_{kn}\frac{1}{2}(\tilde{v}_{kn}^y+\tilde{v}_{k+qn}^y) X_{kn}\dg X_{k+qn},
\eea
with the fermionic, $v_{km}^y = \partial \epsilon_{km}/\partial k_y$ and bosonic, $\tilde{v}_{kn} = \partial \zeta_{kn}/\partial k_y$ velocities.  The diamagnetic contributions are,
\bea
T_F & = & T\sum_{i\omega_n} \sum_{km}\frac{\partial^2 \epsilon_{km}}{\partial k_y^2}f_{km}\dg f_{km}\cr
T_X & = & T\sum_{i\nu_n}\sum_{kn} \frac{\partial^2 \zeta_{kn}}{\partial k_y^2}X_{kn}\dg X_{kn}.
\eea

\figwidth=8.5cm
\fg{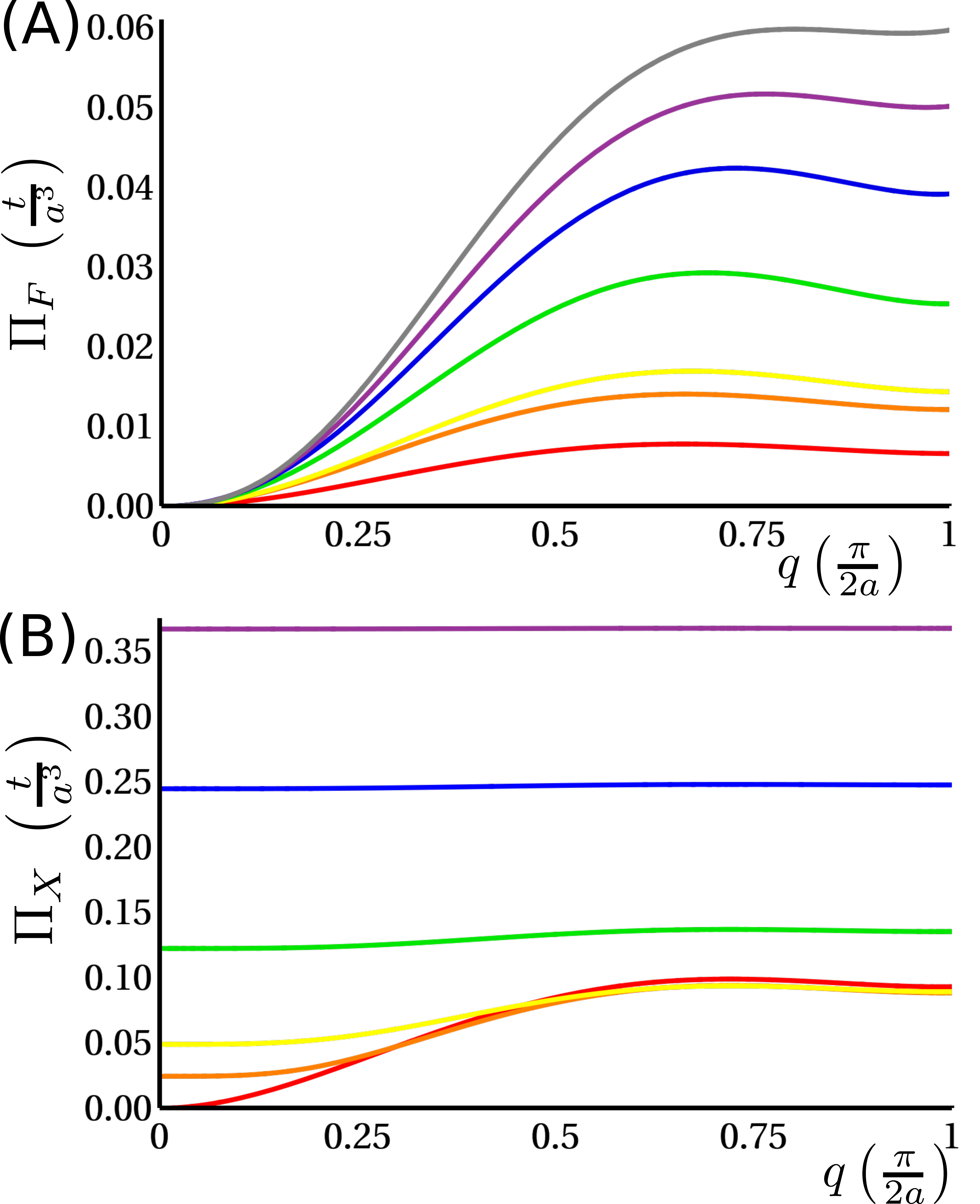}{correlators}{(Color online) (A) Fermionic transverse current-current correlator, $\Pi_F(\bq)$ on the pyrochlore lattice for varying proximity to the Mott transition: $X_0^2 = 0$ (red, $U = 1.4 t$); $X_0^2 = 0.05$ (orange, $U = 1.2 t$); $X_0^2 = .1$ (yellow, $U = 1.1 t$); $X_0^2 = 0.25$ (green, $U = 0.8 t$); $X_0^2 = 0.5$ (blue, $U = .3 t$); $X_0^2 = .75$ (purple, $U=.08t$); and $X_0^2 = 1$ (gray, $U = 0$).  Note that the distance from the Mott transition increases the effective fermionic bandwidth. (B) Bosonic transverse current-current correlator, $\Pi_X(\bq)$ on the pyrochlore lattice for varying proximity to the Mott transition (same color scheme as above).  Note that for large $X_0^2$, the correlator approaches a constant in $q$-space, just as expected in the Fermi liquid.}

The fermionic current-current correlator is shown in Figure 5A for several values of $X_0^2$.  $\Pi_F$ decreases in magnitude with increasing $U$.  

The bosonic correlator requires a bit more care. Deep in the Fermi liquid, the paramagnetic term vanishes, and the diamagnetic term, $T^{\rm FL}_X = -(8/3 t_1 Q_1 + 24 t_2 Q_2) X_0^2$ is the only contribution.  At the QCP, we can calculate $\Pi_X^{QC}(\bq)$ directly.  In between, we must interpolate between the two\cite{Podolsky09},
\begin{equation}
\Pi_X^T(q) = \Pi_B^{QC}(q) \coth \frac{\Pi_B^{QC}(q)}{(8/3 t_1 Q_1 + 24 t_2 Q_2) X_0^2}.
\end{equation}
The results are shown in Figure 5B for a variety of $X_0^2$, and $\Pi_X^T$ generally decreases with increasing $U$, like $\Pi_F^T$.

Having the correlators, we now combine them and Fourier transform,
\bea
J_\kappa^{zz}(\br) & = & -\frac{J_{cf}^2 A^2}{3 \hbar^2}\frac{1}{\Omega} \sum_\bq e^{i \bq \cdot \br}\left[q^2-q^z q^z\right] \frac{1}{\Pi_F^T(\bq) + \Pi_X^T(\bq)}\cr
& = & -\frac{J_{cf}^2 a^7}{8 t \hbar^2} \frac{2}{\pi^3}\int_{BZ} d\tilde{{\bf q}} f(\tilde{q})
\eea
where we have defined $A = \frac{\sqrt{3}a^2}{32}$ as the area of the isosceles triangle in Figure 1B and $a = 10.4\AA$ is the lattice constant.  We have also introduced $\tilde{q} = q a$ and used that the units of $\Pi_{F,X}$ are $t/a^3$ to introduce the dimensionless function $f(\tilde{q}) = t/a^5 (\tilde{q}^2-\tilde{q}_z^2)/[\Pi_F^T(\bq) + \Pi_X^T(\bq)]$.  For $J_{cf}$, we have simply taken $J_K^3 \rho^2$, where $\rho$ is the density of states at the Fermi energy, and we take the estimate $J_K$ from section II.  We took $t = 10^4$K roughly from the density functional theory calculation\cite{Wan11}. 

\fg{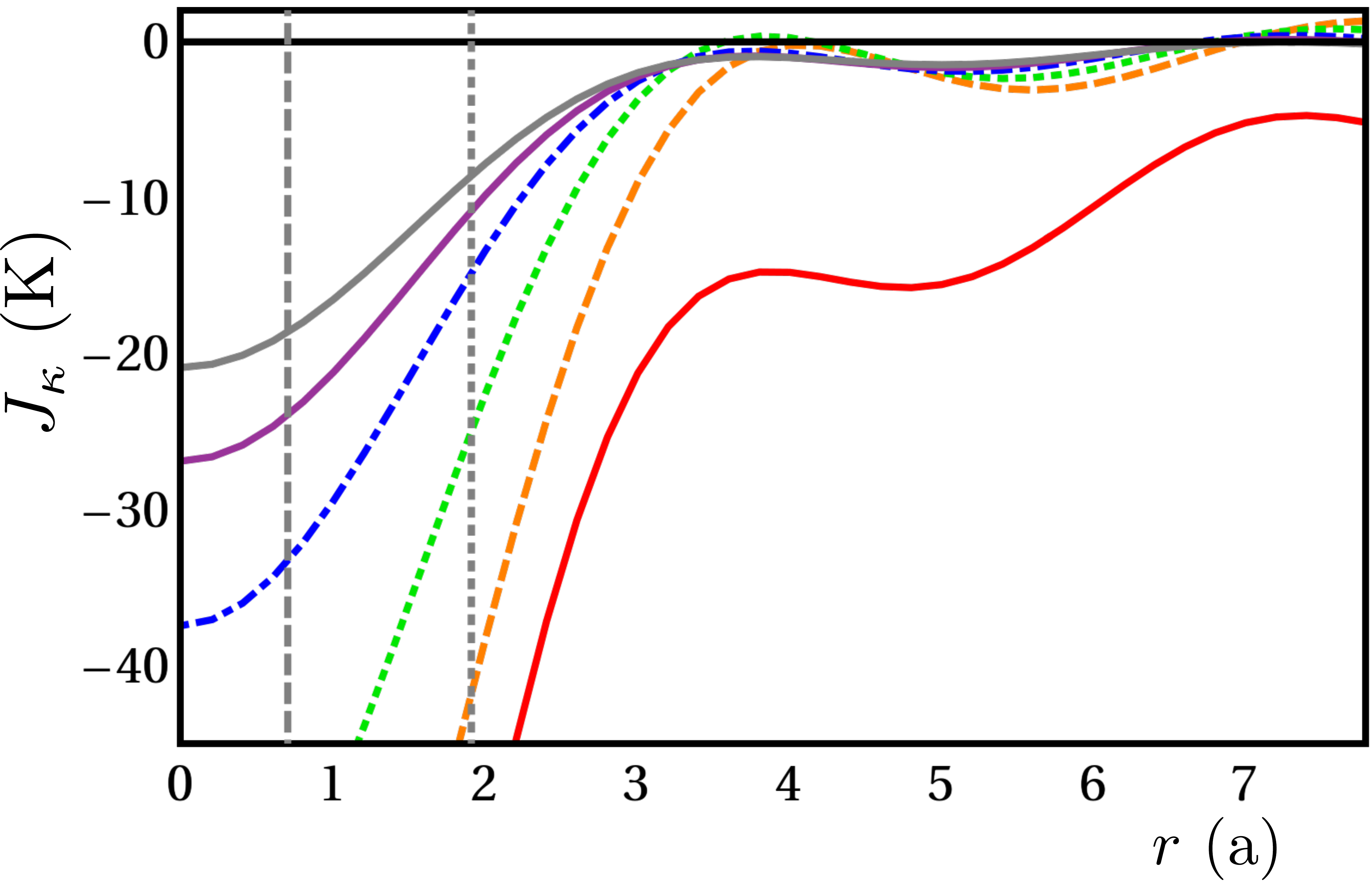}{result}{(Color online) Chiral coupling $J_\kappa(r)$ in real-space for several  $X_0^2$ (color scheme explained in figure 5).  $J_\kappa (r)$ is ferro-chiral for all $r$ and $X_0^2$, and increases in magnitude as the Mott transition is approached.  The coupling constant is given in Kelvin, and even for $U = 0$, the chiral couplings between hexagons, $r = a/\sqrt{2}$ and between layers, $r = \sqrt{11}/{3}a$ are both on the order of 10K.}

The chiral coupling oscillates slowly and is nearly always negative, or ferro-chiral, due to the small carrier density, although for larger distances further from the Mott transition it can be small and antiferro-chiral due to the small pockets. By contrast, on the half-filled fcc lattice, which has a large Fermi surface, the sign of the chiral coupling  is very sensitive to the inter-atomic distance, and other parameters. While the Mott transition does increase the chiral coupling, it is already quite large for free fermions on the pyrochlore lattice, making the chiral RKKY coupling a plausible explanation for \prirop.

\section{Predictions and Speculations}

As \priro differs so substantially from the theoretical systems initially proposed to contain chiral spin liquids, determining the relevant microscopic Hamiltonian is an important problem.  As magnetic interactions seem unlikely to generate the relatively high transition temperature, $T_H = 1.5$K, we propose that the chiral RKKY effect generates a ferrochiral coupling.  Using a slave rotor treatment of the $J_{\rm eff} = 1/2$ pyrochlore lattice to estimate its magnitude and sign, we found that the interaction is always ferrochiral and of the necessary order of magnitude.  The relevant Hamiltonian is then,
\begin{equation}
H = J_1 \sum_{\langle ij \rangle} S^z_i S^z_j + \sum_{ij} J_\kappa(i-j) \kappa_i \kappa_j.
\end{equation}
Studying the ground and excited states of this Hamiltonian is an interesting problem for future work, although unfortunately complicated by the sign problem. One key question is whether or not the observed smallness of the parasitic magnetic moment is consistent with the Hamiltonian and how it might be tuned.

As the chiral RKKY effect requires the presence of conduction electrons, we do not expect the insulating analogues, Pr$_2$Sn$_2$O$_7$[\onlinecite{Zhou08}] and Pr$_2$Zr$_2$O$_7$[\onlinecite{PrZr}] to exhibit chiral order, unlike other theories based on the $\Gamma_3$ Pr doublet\cite{Onoda10}.  Indeed, both these compounds appear to have residual magnetic entropy consistent with spin ice, although the non-chiral nature could be confirmed by RIXS\cite{Ko11}.

\prirop's proximity to a metal insulator transition provides the opportunity to test whether or not driving the system closer to a Mott transition increases the chiral ordering temperature by increasing the conduction electron chiral susceptibility.  The nature of the metal insulator transition is unclear; though its second order nature suggests it has a large Slater character, analogy to other iridates, like Sr$_2$IrO$_4$\cite{Kim08} suggests that it may have some Mott character as well.  As pressure drives the materials further from the MIT, it should suppress the chiral transition in \prirop, while Nd doping should, in principle, increase it.  It is also possible that Nd$_2$Ir$_2$O$_7$ under pressure may realize a chiral state, and Hall effect measurements on single crystals under pressure should be done to check for such a chiral state.

\fg{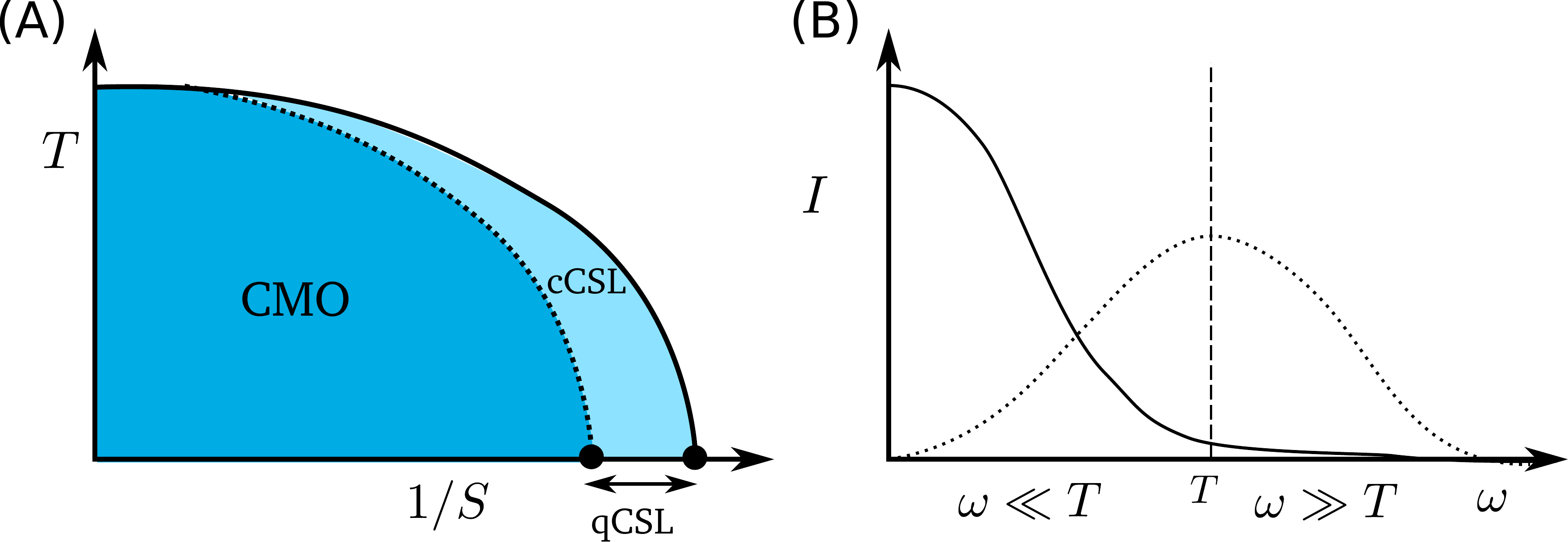}{qvscl}{Classical versus quantum chiral spin liquids.  (A) A generic phase diagram.  The classical CSL* (cCSL*) occurs above a magnetically ordered state with net chirality (CMO), while the quantum CSL* (qCSL*) is only revealed when quantum fluctuations kill the magnetic order. (B) A rough sketch of the frequency dependence of the spin structure factor to illustrate the difference between classical (solid line) spin liquids, whose fluctuations are mainly confined to $\omega < T$ and quantum (dashed line) chiral spin liquids, whose fluctuations have a much broader distribution at finite frequencies.}

An interesting open question is whether the observed CSL* is classical or quantum in nature.  Classical spin liquids are incoherent thermal superpositions of states that occur above magnetically ordered states, while quantum spin liquids involve a coherent superposition of states.  Generically, increasing quantum fluctuations can tune a thermal CSL* into a quantum one (see Figure 7A).  The magnetic quantum criticality seen in the magnetic Gruneisen factor suggests that the ground state is not magnetically ordered, and thus \priro may be a quantum CSL*\cite{Gruneisen}.  This question can be resolved by examining the energy distribution of spin fluctuations, through the frequency dependence of the spin structure factor.  A classical spin liquid is dominated by thermal fluctuations, so most fluctuations occur at $\omega \ll t$, while a quantum spin liquid will have a broader distribution.  Theoretically, quantum CSL*'s may have interesting topological properties: in the 2D case, chiral spin liquids are predicted to have semionic excitations\cite{Wen89} and a quantized thermal Hall effect\cite{thermalhall}, and it is unclear how the three-dimensionality and small parasitic moment will affect the topological nature of the phase.

\emph{Acknowledgements:} The authors would like to acknowledge useful discussions with Leon Balents, Patrick Lee, Ivar Martin, Karen Michaeli, Andrew Millis, Ashvin Vishwanath and Fa Wang.  We acknowledge funding from the
Simons Foundation (RF), NSF DMR-1005434 (TS), and NSF grant 1066293 (RF, TS)
while at the Aspen Center for Physics.  TS was also partially supported
by the Simons Foundation by award number 229736. We are grateful for the
hospitality of the Aspen Center for Physics. TS thanks the Physics
Department at Harvard and the Perimeter Institute for Theoretical
Physics for hospitality where part of this work was done.

%-------------------


\begin{thebibliography}{99}
%-------------------

\bibitem{Nakatsuji06} S. Nakatsuji \emph{et al}, Phys. Rev. Lett. {\bf 96}, 087204 (2006).

\bibitem{Machida10} Y. Machida, S. Nakatsuji, S. Onoda, T. Tayama and T. Sakakibara, Nature {\bf 463}, 210(2010).

\bibitem{Machida05} Y. Machida, S. Nakatsuji, H. Tonomura, T. Tayama, T. Sakakibara, J. van Duijn,C. Broholm, Y. Maeno
J. Phys. Chem. Solids 66, 1435–1437 (2005).

\bibitem{Bramwell01} S. Bramwell and M. Gingras, Science 294, 1495 (2001).

\bibitem{Udagawa12} M. Udagawa, H. Ishizuka and Y. Motome, Phys. Rev. Lett. {\bf 108}, 066406 (2012).

\bibitem{Isakov04} Isakov, S. V., Gregor, K., Moessner, R. and Sondhi, S. L. PRL 93, 167204 (2004).

\bibitem{Kim08} B. J. Kim \emph{et al}, Phys. Rev. Lett. {\bf 101}, 076402 (2008).

\bibitem{Singh10} Yogesh Singh and P. Gegenwart, Phys. Rev. B 82, 064412 (2010).

\bibitem{Yanagashima01} D. Yanagashima and Y. Maeno, J. Phys. Soc. Jpn. 70, 2880 (2001).

\bibitem{Matsuhira07} Kazuyuki Matsuhira {\emph et al}, J. Phys. Soc. Jpn {\bf 76}, 043706 (2007).

\bibitem{Matsuhira11} Kazuyuki Matsuhira, Makoto Wakeshima, Yukio Hinatsu, and Seishi Takagi, J. Phys. Soc. Jpn. {\bf 80}, 094701(2011).

\bibitem{Zhao11} Songrui Zhao. J. M. Mackie, D. E. MacLaughlin, O. O. Bernal, J. J. Ishikawa, Y. Ohta, S. Nakatsuji, Phys. Rev. B (R) (2011).

\bibitem{Wan11} Xiangang Wan, Ari Turner, Ashvin Vishwanath and Sergey Y. Savrasov, Phys. Rev. B {\bf 83} 205101 (2011).

\bibitem{Sakata11} Masafumi Sakata {\emph et al}, Phys. Rev. B {\bf 83}, 041102(R) (2011).

\bibitem{Tafti11} F.F. Tafti, J. Ishikawa, Y. Machida, A. McCollam, S. Nakatsuji, and S.R. Julian (2011).

\bibitem{2CK} See D. L. Cox and A. Zawadowski, “Exotic Kondo effects in metals: magnetic ions in a
crystalline electric field and tunneling centers”, Advances in Physics, 47 599 (1998) and
references therein.

\bibitem{Hastatic} P. Chandra, P. Coleman and R. Flint, arXiv:1207.4828 (2012).

\bibitem{Machida07} Y. Machida, S. Nakatsuji, Y. Maeno, T. Tayama, T. Sakakibara and S. Onoda, Phys. Rev. Lett. {\bf 98}, 057201 (2007).

\bibitem{MacLaughlin09} D.E. MacLaughlin et al, Physica {\bf B 404}, 667 (2009).

\bibitem{Nagaosa10} Naoto Nagaosa, Jairo Sinova, Shigeki Onoda, A. H. MacDonald, and N. P. Ong, Rev. Mod. Phys. {\bf 82}, 1539 (2010).

\bibitem{udegawa12} M. Udagawa and R. Moessner, arXiv:1212.0293 (2012).

\bibitem{moon12}  E.-G. Moon, C. Xu, Y. B. Kim and L. Balents, arXiv:1212.1168 (2012).

% G. A. Fiete et al, Physica E 44, 845 (2012).
%H. Katsura, N. Nagaosa, and P.A. Lee, PRL {\bf 104}, 066403 (2010).

% References to find:
% NdMo reference (can do now - DONE)
% basic PrZr reference (internet - DONE)
% TbTi (internet - DONE)
% Sr213 (internet - mean Na2IrO3)
% Ko (internet- DONE)
% background gauge field - see Ilya's thesis (now - DONE)
% what's a basic 2CK/criticality reference (now - DONE)
% thermal hall reference (internet - DONE)
% Do I need more references?  Possibly more for Mott transition


% Below here need to cite all of these
% Really need to sort out how to cite the Mott transition
% What did Gang Chen do again, and is it published?
% will cite near something about chiral RKKY being sixth order

%\bibitem{Chen11} Gang Chen - not published

%\bibitem{Coleman83} P. Coleman, Phys. Rev. B 28, 5255 (1983). 

%\bibitem{balents02} L. Balents, M.P.A. Fisher and S. Girvin, {\bf PRB} 65, 224412 (2002).

%\bibitem{Libby91} S.B. Libby, Z. Zou and R.B. Laughlin, Nucl. Phys. B {\bf 348}, 693(1991).

\bibitem{Ko11} Wing-ho Ko and Patrick Lee, Phys. Rev. B 84, 125102 (2011).

\bibitem{Kalmeyer87} V. Kalmeyer and R. Laughlin, PRL {\bf 59}, 2095 (1987).

\bibitem{Wen89} X.G. Wen, Frank Wilczek and A. Zee, Phys. Rev. B {\bf 39}, 11413(1989).

\bibitem{RKKY}M.A. Ruderman and C. Kittel, Phys. Rev. 96, 99(1954); T. Kasuya, Prog. Theor. Phys. 16, 45 (1956);K. Yosida, Phys. Rev. 106, 893 (1957).

\bibitem{Ikeda08} A. Ikeda and H. Kawamura, JPSJ 77, 073707 (2008).

\bibitem{Hermele04} M. Hermele, M. P. A. Fisher, L. Balents PRB 69, 064404 (2004).

\bibitem{Chen12} G. Chen and M. Hermele, arXiv:1208.4853 (2012).

\bibitem{TbTi} J. S. Gardner et al, Phys. Rev. Lett. {\bf 82}, 1012 (1999). 

\bibitem{TbTiTheory} Lucile Savary and Leon Balents, Phys. Rev. Lett. {\bf 108}, 037202 (2012).

\bibitem{backgroundgauge} K. Ohgushi, S. Murakkami, and N. Nagaosa, Phys. Rev. B 62, R6065 (2000).

\bibitem{Florens04} S. Florens and A. Georges PRB 70, 035114 (2004).

\bibitem{Lee92} Patrick A. Lee and Naota Nagaosa, Phys. Rev. B {\bf 46}, 5621 (1992).cdl

%\bibitem{Ioffe89} L. Ioffe and A. Larkin, Phys. Rev. B {\bf 39}, 8988 (1989).

\bibitem{Pesin10} Dmytro Pesin and Leon Balents, Nature Physics {\bf 6}, 376(2010).

\bibitem{Witczak12} W. Witczak-Krempa and Y.B. Kim, Phys. Rev. B 85, 045124 (2012).

\bibitem{Kargarian11} M. Kargarian, J. Wen, and G. A. Fiete, PRB 83, 165112 (2011).

\bibitem{Podolsky09} Daniel Podolsky, Arun Paramekanti, Yong Baek Kim and T. Senthil, Phys. Rev. Lett. {\bf 102}, 186401 (2009).

\bibitem{Onoda10} Shigeki Onoda and Yoichi Tanaka, PRL {\bf 105}, 047201 (2010).

\bibitem{Gruneisen} P. Gegenwart and S. Nakatsuji - unpublished.

\bibitem{thermalhall} C.L. Kane and M.P.A. Fisher, PRB {\bf 55}, 15832  (1997).

\bibitem{ndmo} Y. Taguchi, Y. Oohara, H. Yoshizawa, N. Nagaosa, and Y.Tokura, Science 291, 2573 (2001).

\bibitem{Zhou08} H.D. Zhou et al, PRL {\bf 101}, 227204 (2008).

%\bibitem{Burnell} For details of the pyrochlore space group operators, see F. J. Burnell, Shoibal Chakravarty, and S. L. Sondhi, Phys. Rev. B 79, 144432 (2009).

%\bibitem{Lee12} S.B Lee, S. Onoda and L. Balents, arxiv:1204.2262v2 (2012).

%\bibitem{Lee05} Sung-Sik Lee and Patrick A. Lee, Phys. Rev. Lett. 95, 036403 (2005).

\bibitem{PrZr} K. Matsuhira et al, J. Phys.: Conf. Ser. 145 012031 (2009).

\bibitem{Balicas11} L. Balicas, S. Nakatsuji, Y. Machida and S. Onoda, Phys. Rev. Lett. {\bf 106}, 217204 (2011).

\end{thebibliography}
\end{document}